\documentclass[pdflatex,iicol,sn-mathphys-num]{sn-jnl}
\usepackage{graphicx}%
\usepackage{multirow}%
\usepackage{amsmath,amssymb,amsfonts}%
\usepackage{amsthm}%
\usepackage{rsfso}%
\usepackage[title]{appendix}%
\usepackage{xcolor}%
\usepackage{textcomp}%
\usepackage{manyfoot}%
\usepackage{booktabs}%
\usepackage{listings}%
\usepackage{xspace}
\usepackage{siunitx}

\let\orcidlogo\relax
\usepackage{orcidlink}

\newcommand{\footnoteA}{\footnote}

\newcommand{\gm}{\ensuremath{g\!-\!2}\xspace}
\newcommand{\gmtwo}{\gm}

\newcommand{\RunOne}{Run-1\xspace}
\newcommand{\RunTwo}{Run-2\xspace}
\newcommand{\RunThree}{Run-3\xspace}
\newcommand{\RunFour}{Run-4\xspace}
\newcommand{\RunFive}{Run-5\xspace}
\newcommand{\RunSix}{Run-6\xspace}

\newcommand{\authornote}[1]{{%
  \let\thempfn\relax
  \footnotetext[0]{$\diamond${ }#1}
}}

\usepackage{lineno}

\makeatletter

\renewcommand{\Authorfont}{%
  \reset@font\fontsize{9.5bp}{11.3bp}\selectfont
  \titraggedcenter
}
\renewcommand{\addressfont}{%
  \reset@font\fontsize{8.5bp}{10.2bp}\selectfont
  \titraggedcenter
}

\let\gTwoOriginalMakeTitle\maketitle
\renewcommand{\maketitle}{%
  \onecolumn
  \raggedbottom
  \begingroup
    \providecommand{\thefootnoteA}{}%
    \renewcommand{\thefootnoteA}{}%
    \@ifundefined{footnotetextA}{%
    \long\def\footnoteA##1{%
        \begingroup
        \renewcommand{\thefootnote}{}%
        \footnotetext[0]{##1}%
        \endgroup
    }%
    }{%
    \let\footnoteA\footnotetextA
    }%
    \gTwoOriginalMakeTitle
  \endgroup
  \twocolumn
  \flushbottom
}

\makeatother

\title{Data handling, production, and simulation in the Muon \texorpdfstring{$g-2$}{g-2} Experiment at Fermilab}

\makeatletter
\providecommand{\sncollaboration}[1]{%
  \g@addto@macro\artauthors{%
    \par\addvspace{5pt}%
    {\normalfont\fontsize{10bp}{12bp}\selectfont(#1)}\par
  }%
}
\makeatother

\newif\ifgTwoFirstAlternateNote
\gTwoFirstAlternateNotetrue

\providecommand{\snaltaffnote}[2]{%
  \par\noindent\textsuperscript{#1}\,#2\par%
}

\author[32]{D.~P.~Aguillard\orcidlink{0000-0002-5556-1169}}
\author[29]{T.~Albahri\orcidlink{0000-0002-1191-5463}}
\author[7]{D.~Allspach\orcidlink{0009-0008-2417-5338}}
\author[7]{J.~Annala}
\author[7]{K.~Badgley\orcidlink{0000-0002-3942-5108}}
\author[34]{S.~Bae{\ss}ler\orcidlink{0000-0001-7732-9873}}
\author[26]{L.~Bailey\orcidlink{0009-0004-7290-8963}}
\author[27,a]{E.~Barlas-Yucel\orcidlink{0000-0001-9562-2125}}
\author[6]{T.~Barrett\orcidlink{0000-0002-0807-5846}}
\author[7]{E.~Barzi\orcidlink{0000-0001-5829-2147}}
% ...
\sncollaboration{The Muon \gmtwo Collaboration}

\affil[1]{Argonne National Laboratory, Lemont, Illinois, USA}
% ...
\affil[5]{Center for Axion and Precision Physics (CAPP) / Institute for Basic Science (IBS)}
\affil[6]{Cornell University, Ithaca, New York, USA}
\affil[7]{Fermi National Accelerator Laboratory, Batavia, Illinois, USA}
% ...
\affil[26]{Department of Physics and Astronomy, University College London, London, United Kingdom}
\affil[27]{University of Illinois at Urbana-Champaign, Urbana, Illinois, USA}
% ...
\affil[29]{University of Liverpool, Liverpool, United Kingdom}
% ...
\affil[32]{University of Michigan, Ann Arbor, Michigan, USA}
% ...
\affil[34]{University of Virginia, Charlottesville, Virginia, USA}
% ...

\artnote{%
  \begingroup
  \setlength{\parindent}{0pt}%
  \setlength{\parskip}{1pt}%
  \snaltaffnote{a}{Now at Fermi National Accelerator Laboratory, Batavia, Illinois, USA.}%
  \snaltaffnote{b}{Also at Universit\`a di Trieste, Trieste, Italy.}%
  \snaltaffnote{c}{Also at INFN Gruppo Collegato di Udine, Sezione di Trieste, Udine, Italy.}%
  \snaltaffnote{d}{Also at State Key Laboratory of Dark Matter Physics, Shanghai, China; also at Key Laboratory for Particle Astrophysics and Cosmology (MOE), Shanghai, China; also at Shanghai Key Laboratory for Particle Physics and Cosmology, Shanghai, China.}%
  \snaltaffnote{e}{Also at Universit\`a di Napoli, Naples, Italy.}%
  \snaltaffnote{f}{Also at Lebedev Physical Institute, Moscow, Russia; also at National Research Nuclear University MEPhI, Moscow, Russia.}%
  \snaltaffnote{g}{Also at Universit\`a di Pisa, Pisa, Italy.}%
  \snaltaffnote{h}{Also at Istituto Nazionale di Ottica - Consiglio Nazionale delle Ricerche, Pisa, Italy.}%
  \snaltaffnote{i}{Now at Amherst College, Amherst, Massachusetts, USA.}%
  \snaltaffnote{j}{Now at Department of Physics and Astronomy, Trinity University, San Antonio, Texas, USA.}%
  \snaltaffnote{k}{Now at Alliance University, Bangalore, India.}%
  \snaltaffnote{l}{Also at INFN, Sezione di Roma Tor Vergata, Rome, Italy.}%
  \snaltaffnote{m}{Now at INFN, Laboratori Nazionali di Frascati, Frascati, Italy.}%
  \snaltaffnote{n}{Now at University of Washington, Seattle, Washington, USA.}%
  \snaltaffnote{o}{Now at Istinye University, Istanbul, T\"urkiye.}%
  \snaltaffnote{p}{Also at University of Rijeka, Rijeka, Croatia.}%
  \snaltaffnote{q}{Now at Department of Physics, Xiamen University Malaysia, Sepang, Selangor, Malaysia.}%
  \snaltaffnote{r}{Also at Zhejiang Lab, Hangzhou, Zhejiang, China.}%
  \snaltaffnote{s}{Also at Shenzhen Technology University, Shenzhen, Guangdong, China.}%
  \snaltaffnote{t}{Also at Novosibirsk State University, Novosibirsk, Russia.}%
  \snaltaffnote{u}{Also at Scuola Normale Superiore, Pisa, Italy.}%
  \snaltaffnote{v}{Also at Northeastern University, Boston, Massachusetts, USA.}%
  \snaltaffnote{w}{Now at Northwestern University, Evanston, Illinois, USA.}%
  \snaltaffnote{x}{Now at Virginia Tech, Blacksburg, Virginia, USA.}%
  \snaltaffnote{y}{Now at TU Dortmund University, Dortmund, Germany.}%
  \snaltaffnote{z}{Now at Cornell University, Ithaca, New York, USA.}%
  \snaltaffnote{aa}{Now at Innovative Solutions R\&D LLC, Stony Brook, New York, USA.}%
  \snaltaffnote{ab}{Also at Universit\`a di Roma Tor Vergata, Rome, Italy.}%
  \snaltaffnote{ac}{Also at University of Hawaii at Manoa, Honolulu, Hawaii, USA.}%
  \snaltaffnote{ad}{Also at INFN, Sezione di Pisa, Pisa, Italy.}%
  \snaltaffnote{ae}{Now at Boston University, Boston, Massachusetts, USA.}%
  \snaltaffnote{\ensuremath{\dagger}}{Deceased.}%
  \endgroup
}

\date{\today}

\abstract{
This paper details the computing strategies of the Muon \gmtwo Experiment at Fermilab, which measured the muon's anomalous magnetic moment with a precision of 127 ppb by processing over 10 petabytes of detector and simulation data. This paper discusses the computational and operational challenges of data production, data handling, and data analysis of such large datasets. These challenges were overcome by using Fermilab-developed software infrastructure and production tools, making efficient use of resources, and improving the collaboration-developed reconstruction workflow. By overcoming these challenges, the experiment improved the data throughput capacity and reduced the analysis turnaround time by an order of magnitude over the course of its lifetime.}

\begin{document}

\maketitle


\section{Introduction}

The Muon \gmtwo Experiment conducted a multiyear research program in particle physics hosted at Fermilab\cite{Muong-2:2024hpx, PRL_Run-4/5/6}. During beam data collection, the Muon \gmtwo Experiment accumulated about 15,000 files per day, each one 2\,GB in size.
Over the course of six years of data taking we collected approximately 10\,PB of data corresponding to about $10^7$ files. In addition, physics simulations produced about 2\,PB of data.
The role of production is to perform the physics reconstruction of all the acquired data and select good quality data. The reconstruction process needed to be highly reproducible and capable of running multiple reconstruction algorithms so that several independent analyses could be performed for cross checks and systematic uncertainty estimations. Each file had to be processed at least twice: first, a pre-production stage determined data-dependent calibrations, and then the full production stage performed the reconstruction. Data quality was assessed and reconstruction performance was validated throughout the processing chain. Given that all the data had to be reconstructed and analyzed, the speed and efficiency of the reconstruction process was of crucial importance and represented a significant computing challenge.\\

In this paper, we describe how the collaboration successfully performed this computing task. Section \ref{sec:comp_overview} will provide an overview of Muon \gmtwo computing and the overall computing strategy adopted. Section \ref{sec:comp_infrastructure} will describe the hardware and software frameworks provided by Fermilab that are used by multiple experiments. Sections \ref{sec:production}, \ref{sec:analysis} and \ref{sec:simulation} will describe how these framework tools are used to address the Muon \gmtwo production reconstruction, simulation, and analysis objectives. Finally, Section \ref{sec:timeline} will discuss the improvements that were made in our computing over the course of the experiment. Section \ref{sec:conclusion} offers some conclusions and lessons learned.

\section{Muon \gmtwo Computing Strategy}\label{sec:comp_overview}


The main goal of the Muon \gmtwo computing infrastructure was to support analysis workflows that extracted physics results from processed data. Calorimeter data were analyzed to extract the muon magnetic anomaly and tracker data were analyzed to understand the beam dynamics effects that can affect the results. A description of the \gmtwo apparatus can be found in Ref.~\cite{Muong-2:2024hpx} and references therein. The extracted physics results had to be precise, robust, and reproducible while also making efficient use of resources (both computing and personnel). The \gmtwo computing model encompassed two production workflows: experimental data processing (Sec.~\ref{sec:comp_infrastructure}-\ref{sec:analysis}) and Monte Carlo (MC) simulation (Sec.~\ref{sec:simulation}). \\

Experimental data processing began with the raw detector data and organized it into discrete events corresponding to muon fills or magnetic field measurement periods, with each event serving as a processing unit. Upon collection, raw data were first stored on Fermilab's shared computing infrastructure and archived to tape storage. This \underline{file transfer system} had to be robust to failures such as accidental file deletion and duplication.  \\

Subsequently, the stored data were processed through multiple production stages that progressively transformed raw detector waveforms into calibrated, physics analysis-ready data products (Sec.~\ref{sec:production} and Sec.~\ref{sec:analysis}). This processing began with minimal reconstruction to extract calibration constants using multiple independent reconstruction algorithms and was followed by complete detector reconstruction that applied these calibrations, ultimately producing the high-level data products used in the analysis workflows. This required an \underline{event-processing framework} and \underline{conditions database} that could handle changes in calibration constants throughout the lifetime of the experiment. \\

The precision targeted by the analyses required the minimization of both statistical and systematic errors. Lowering the statistical error required analyzing every detected positron and evaluating the systematic error required that the full analysis be performed multiple times using a variety of techniques. To ensure that the physics results were robust, there were four independent calorimeter data reconstruction methods~\cite{Aaron, Sweigart, Girotti, Kelton} that transformed the raw waveform data from calorimeter modules into deposited energies and times. Data from the tracker stations were also reconstructed. For this to be reproducible we needed a \underline{job submission system} that could submit jobs to the grid and automatically recover failed jobs to ensure that all data were successfully processed every iteration. \\



The Muon \gmtwo computing model had to optimize the use of both limited computing and personnel resources.  Limited human resources strongly encouraged the use of common infrastructure services developed by Fermilab computing experts and intended for use by multiple experiments (Sec~\ref{sec:comp_infrastructure}).
Furthermore, data processing was organized into a set of production steps so that all members of the experiment had easy access to different stages of data, and the full collaboration signed up for \lq\lq offline production shifts\rq\rq\ to manage the stages of data production based on guidance from experiment production experts.  The entire process was fully documented and explained in detail to all collaborators, and the code was version-controlled for reproducibility. This meant all the systems used by shifters had to be \underline{user friendly}. Resource limitations encouraged the use of \underline{grid resources} for processing and efficiently sharing both production and storage resources among multiple experiments based on annual presentations of overall computing needs. \\ 

During the course of the experimental data taking we went through six different running periods, roughly one per year. Each successive running period included improved experimental conditions, larger numbers of events, greater sophistication of analysis techniques, and additional calibration and sensitivity measurements, to provide a continual reduction in both statistical and systematic errors.  Production strategies evolved over the experiment lifetime to cope with the order of magnitude increases in the amount of data and to preserve timely publication schedules.  To benefit from the increase in data, a greater efficiency was required: this was made possible by more efficient resource usage, increased management and organization of production activities, collaboration-wide agreements on definitions of skimmed datasets (to allow the larger numbers of events in later datasets to be preserved online in roughly constant disk space), and, for the final year of data taking, running data pre-production in a near real-time basis so that we were able to start full production much earlier than in previous runs.  \\

\section{Computing infrastructure}\label{sec:comp_infrastructure}

\subsection{Fermilab Computing Environment and Services}
\label{sec:fnal-computing-services}

The Muon \gmtwo Experiment at Fermilab required a robust and scalable computing infrastructure to process large volumes of detector and simulation data efficiently. The infrastructure supported the full data life cycle, from raw acquisition and archival storage to calibration extraction, reconstruction, validation, skimming, and final analysis.

The computing model of the Muon \gmtwo Experiment was built on Fermilab-developed infrastructure and production tools, including FabrIc for Frontier Experiments (FIFE)~\cite{Herner:2017SF}. FIFE provides a common service layer used by multiple Fermilab experiments, reducing duplication of effort and providing a consistent interface to storage, data handling, grid submission, monitoring, and production management.

The lowest layer of the infrastructure consists of dCache~\cite{10.5281/zenodo.23085911} disk pools, long-term tape storage, and database services. In this section, a \textit{pool} denotes a disk-backed staging or cache area rather than a processing algorithm. The \textbf{GM2 Write Pool} temporarily buffers newly acquired MIDAS raw-data files, denoted by the \texttt{.mid} extension, before archival to tape. The \textbf{GM2 Read Pool} stages raw files recalled from tape for large-scale production, while the \textbf{General Read/Write Pool} holds reconstructed and simulated outputs for downstream processing and analysis. Conditions, calibration, simulation, and data-quality information are stored in database services that are accessed during reconstruction and validation.

The Muon \gmtwo Experiment employed a distributed computing model to provide the required processing capacity. This model used resources from FermiGrid, the Open Science Grid (OSG)~\cite{osg07,osg09}, and the National Energy Research Scientific Computing Center (NERSC), which allowed CPU slots to be allocated dynamically according to production and simulation needs. The scalability of the workflow followed from the parallel nature of Muon \gmtwo offline jobs: each job processed an independent set of input files, typically using a single CPU core and memory below the per-slot limit, and produced a corresponding set of output files. This structure allowed thousands of jobs to run concurrently, provided that the storage and output-throughput requirements were also satisfied.

On top of these raw infrastructure services, the experiment used SAM (Sequential Access via Metadata)~\cite{SAM_2014} to define datasets, track file locations, and provide metadata-driven access to files. The Intensity Frontier Data Handling (IFDH) tools~\cite{Lyon:2014JP} staged input files to worker nodes, while the File Transfer Service (FTS) cataloged and archived output files according to configured rules. Production campaigns were submitted, monitored, and recovered through FIFE job-submission tools and the POMS tool~\cite{POMS_2020}.

The data processing layer running on the worker nodes was the \textit{art} framework~\cite{art_2012}. In the Muon \gmtwo workflow, \textit{art} executed experiment-specific reconstruction, calibration, validation, and simulation modules and stored the outputs as ROOT files with typed data products and provenance information. The resulting production files were commonly referred to as \textit{art}-ROOT files, or \texttt{.art} files in the workflow diagram. ROOT provided the underlying object I/O and was also the standard lightweight analysis environment. After production and skimming, selected reconstructed quantities were converted into standalone \texttt{.root} files containing trees and histograms, which are more convenient for high-level physics analyses that do not require the full \textit{art} environment.

\begin{figure*}[!ht]
    \centering
    \includegraphics[width=\textwidth]{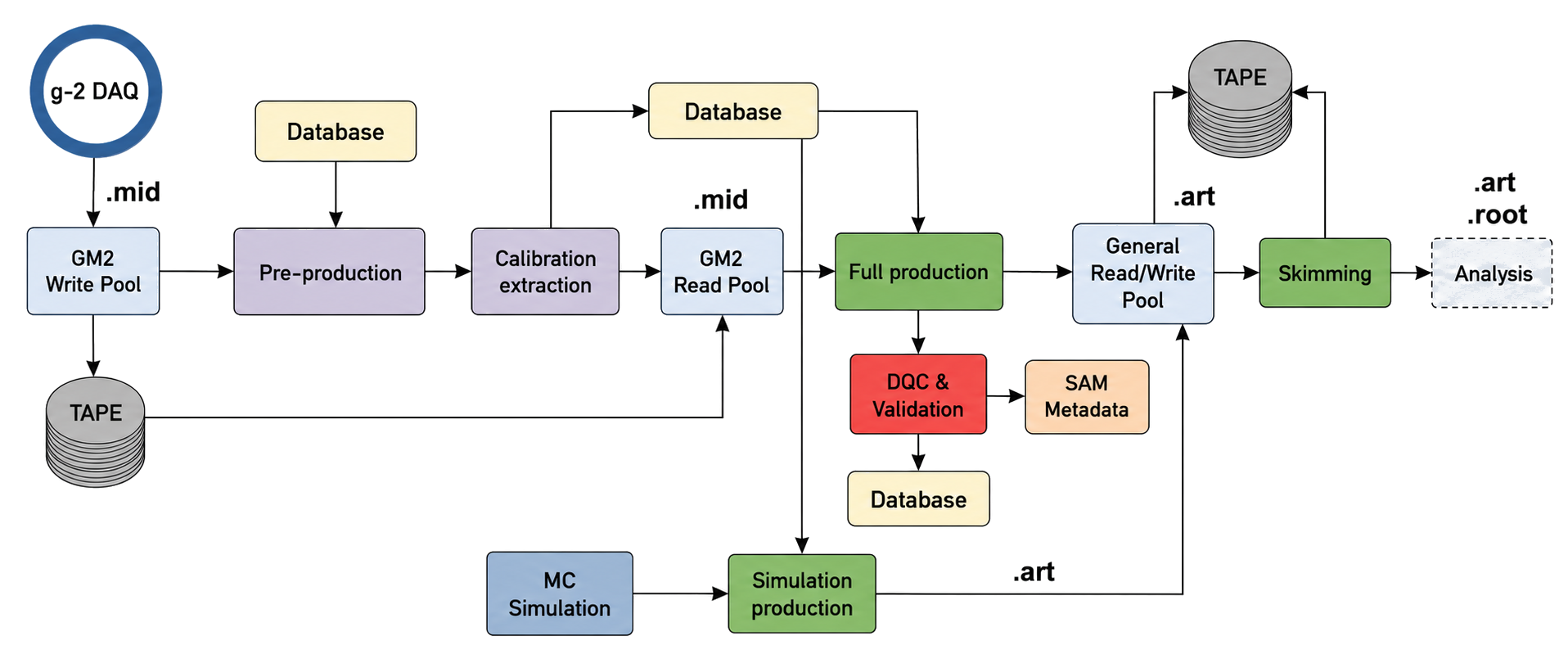}
    \caption{Overview of the Muon \gmtwo computing infrastructure. In the experimental data path (upper branch), the DAQ writes raw \texttt{.mid} files to the GM2 Write Pool, a temporary disk buffer, and the files are archived to tape. Pre-production retrieves configuration and calibration information from the database and performs an initial processing pass to prepare the data for calibration extraction. The extracted constants are written to the central conditions database. For full production, raw files are recalled from tape into the GM2 Read Pool and reconstructed using constants from this database. Full production \texttt{.art} files are written to the General Read/Write Pool, a shared disk staging area, and archived to tape. In parallel, DQC \& Validation stores quality results in a dedicated database and updates the SAM metadata catalog. Skimming produces reduced \texttt{.art} or \texttt{.root} datasets for analysis, with skimmed outputs archived as needed. In the MC path (lower branch), MC Simulation provides generated events to Simulation production, which uses conditions from the central database and writes \texttt{.art} files to the General Read/Write Pool; from there, simulated and experimental samples share the same storage, skimming, and analysis path. Blue boxes denote data sources and disk storage pools; lavender boxes denote pre-processing and calibration; green boxes denote production or data-reduction stages; red denotes data-quality validation; yellow denotes database services; orange denotes the SAM metadata catalog; gray cylinders denote long-term tape storage; and the dashed white box denotes the user-analysis stage. The \texttt{.mid} extension identifies raw MIDAS files, \texttt{.art} denotes \textit{art}-ROOT production files containing \textit{art} data products and provenance, and \texttt{.root} denotes ROOT analysis files.}
    \label{fig:overview}
\end{figure*}

\begin{figure*}[t]
    \centering
    \includegraphics[width=0.9\textwidth]{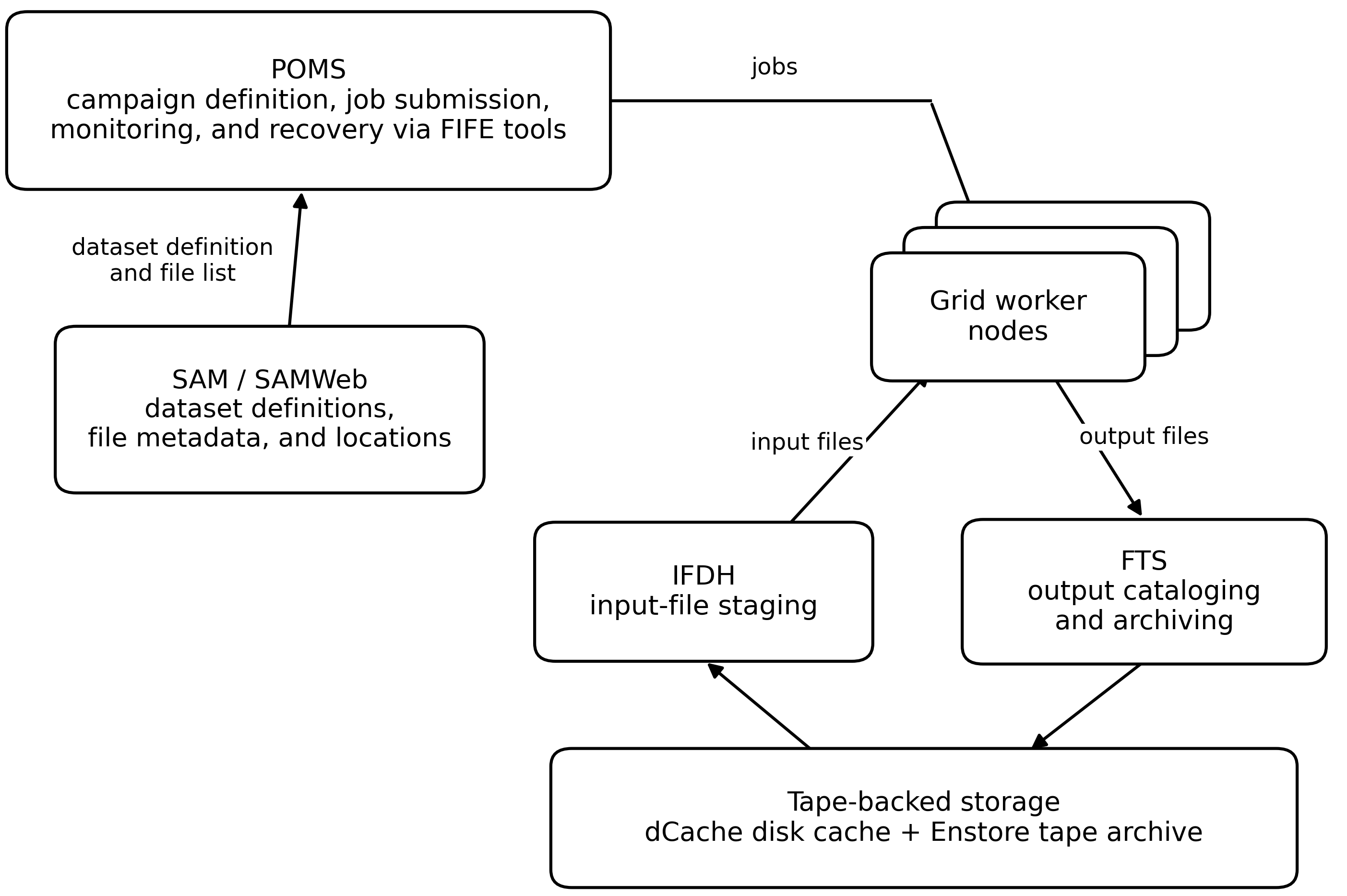}
    \caption{Diagram of the Fermilab-developed infrastructure and how it was used in the Muon \gmtwo experiment.
    Muon \gmtwo collaboration configures a campaign in POMS, while SAM provides the dataset definition, file
    metadata, and file locations. POMS uses FIFE job-submission tools to
    submit, schedule and monitor jobs on grid worker nodes. IFDH stages input files from tape-backed storage to the worker nodes. Output files are passed to FTS, which catalogs them, archives them through the dCache~\cite{10.5281/zenodo.23085911} and Enstore~\cite{osti_15016939}
    storage systems, and registers the corresponding metadata in SAM. Note that Fermilab replaced the Enstore system with the CERN Tape Archive (CTA) system~\cite{Davis:2024nqw} in May 2025.
    }
    \label{fig:fnal-infra}
\end{figure*}

Figure~\ref{fig:overview} gives a overview of the Muon \gmtwo computing infrastructure. Figure~\ref{fig:fnal-infra} complements it by showing
how the Fermilab-developed services were used by the experiment.
Together, two figures connect the experiment's computing workflow stages to the underlying storage systems, metadata services, and job management tools.

\subsubsection{Data Processing Framework}
The \textit{art} data processing framework~\cite{art_2012} was used to process both offline and simulation data on an event-by-event basis, where for most \gmtwo data, an event corresponds to a muon fill. The \textit{art} framework comprises libraries and runtime-configurable main programs, all written in the C++ language. There is a clear boundary between the framework and the user code. The former is developed and maintained by Fermilab software engineers specializing in particle physics infrastructure software. The latter is developed by Muon \gmtwo physicists to define analysis modules that process experimental and simulation data in a coordinated workflow.

In addition to its event-handling capabilities, the use of the \textit{art} framework facilitates a systematic approach to data management through the concept of ``data products.'' In \textit{art} terminology, a data product is an output object produced by a reconstruction, calibration, validation, or simulation module, for example a fitted pulse, a reconstructed calorimeter cluster, or a reconstructed track in the tracker. These objects are stored together with metadata that identify the producing module and process, thereby preserving the provenance of each processing stage. This structure allows physicists to trace how reconstructed quantities were produced, including the transformations, calibrations, and selections applied at each step. Such traceability is essential for validating data integrity, comparing independent reconstruction methods, and refining the accuracy of the physics measurements derived from these events.

\subsubsection{Database Applications}
\label{sec-database}
The Muon \gmtwo Experiment used a Fermilab conditions database system, originally developed for MINERvA~\cite{Aliaga2014}, to manage both data-calibration and simulation constants. The \gmtwo conditions database organized constants by schemas that specify the constant data types and purposes. These constants were organized into folders, each folder serving a specific calibration or configuration purpose, for example detector gain constants, timing offsets, or magnetic-field harmonics. Within each folder, data were stored as time-segmented entries, each tagged with an ``interval of validity'' (IOV) that defined the time range or run period over which the constants were to be applied. Although the numerical values differed from one IOV to another, the underlying database schema, namely the structure and field definitions used to describe the constants, remained identical across all IOVs. This uniform schema ensured that the same software interface can access and interpret data across different validity intervals without modification. Additionally, constants could be tagged for further categorization. Constants were systematically arranged by ``channel'', where each channel, identified by a long integer starting from 1, represented a specific detector element such as a calorimeter crystal. For global values applicable across multiple channels, a separate single-channel folder was used to optimize database storage. 

When reading the conditions database, customized tools verified the IOV, thereby avoiding the retrieval of incorrect constants. The production workflow used a ``data product-producing service'' in \textit{art} to read constants from the database at the start of each subrun. These constants were then transferred into a data product, which was subsequently used within the analysis modules, bypassing direct database access. This method functioned as an internal caching mechanism, ensuring that constants were not retrieved multiple times for the same IOV, thereby enhancing safety and efficiency by reducing repeated database queries for constants.

\subsection{Data Handling and Management}
\label{sec:data-handling-management}

Data handling is the layer that connects file locations, metadata, transfer services, and grid execution. Raw data from the Muon \gmtwo DAQ were initially stored in the \textbf{GM2 Write Pool}, a dedicated disk storage area of about 60~TB designed for rapid data ingestion. This temporary storage provided prompt access for early processing steps before the data were transferred to long-term tape storage.

Following initial storage and archival, the data underwent pre-production processing, during which calibration constants were extracted. After calibration, the data were staged or moved to the \textbf{GM2 Read Pool}, a dedicated storage area optimized for transferring raw files to full-production jobs. The processed data were subsequently validated through the Data Quality Control (DQC) process and recorded in the experiment's database.

After successful production, the data were made available in the \textbf{General Read/Write Pool}, a shared disk storage area among most experiments at Fermilab. At this stage, additional processing, such as skimming, was conducted to extract relevant event subsets. The final data products, stored in \texttt{.art} and \texttt{.root} formats, were then used for physics analysis. Output files were organized in the Fermilab \textit{pnfs} storage namespace, which provided a filesystem-like interface to the experiment's centrally managed disk and tape-backed storage areas. In practice, production workflows limited each \textit{pnfs} directory to about 5000 files. This was an operational limit rather than a data-integrity limit: exceeding it can make directory listings and metadata operations slow or unreliable for production workflows, so files were distributed across multiple directories.

The Muon \gmtwo Experiment generated approximately 9~PB of raw data, with the size of the reconstructed data being about the same. In addition, simulation data occupied around 2-3~PB. Given this data volume and the large fraction stored on tape, efficient data handling and management mechanisms were required. SAM provides the metadata catalog that connected dataset definitions to physical file locations across disk and tape storage. Through SAMWeb interfaces, users and production jobs queried metadata, defined datasets, and obtained file-delivery information without needing to know where each file was physically located.

SAM was originally developed at Fermilab and used by the D0 and CDF experiments~\cite{D0_SAM,CDF_SAM}. It was later modernized to use a lightweight interface~\cite{SAM_2014}. SAM indexes and manages file locations using experiment-specific metadata, enabling file access through metadata-driven queries and allowing the definition of datasets. During production, multiple worker nodes can connect to a SAM project, which consisted of a collection of files that met a specific metadata query, with SAM providing access through URL links. Using the IFDH tools~\cite{Lyon:2014JP}, worker nodes retrieved files via these URLs. This setup streamlined job execution on external grids by abstracting the complexities of data sourcing and transfer protocols, ensuring seamless access to the required data without requiring detailed knowledge of its origin or the transfer protocol involved.

\subsubsection{Data Retrieval and Prestaging}
\label{sec:prestaging}

The SAM framework enabled the efficient distribution of files across multiple worker nodes. However, accessing a large volume of data stored on tape can significantly slow down data retrieval. To mitigate this, traditional prestaging transfers files from tape storage to disk storage in anticipation of their use in grid jobs, usually at the dataset level. This approach is inefficient when the files in a dataset are distributed over many tapes or when only a small number of files are needed from each tape. Frequent tape mounts and dismounts introduce substantial overhead, because each access requires the tape to be loaded, positioned, read, and unloaded. This frequent mechanical handling not only prolongs the data retrieval process but also increases wear and tear on the tape hardware, compounding the inefficiencies associated with small, discrete data access operations. Given the shared tape storage system at Fermilab and its limited throughput, prestaging must therefore be planned carefully to support seamless offline data processing.

By the final production campaigns, the Muon \gmtwo Experiment predominantly used \textbf{migration mode} for prestaging. Unlike traditional dataset-level prestaging, migration mode operates at the tape level by transferring all files from selected tapes to disk. This reduces repeated tape handling and improves throughput when most of the files on the selected tapes are needed. The tradeoff is disk space efficiency: migration mode can stage files that are not immediately needed if only a small fraction of a tape is required. For large production campaigns, however, this tradeoff is favorable because it minimizes tape mounts and improves the overall data-processing rate.

\begin{figure*}[t]
    \centering
    \includegraphics[width=\textwidth]{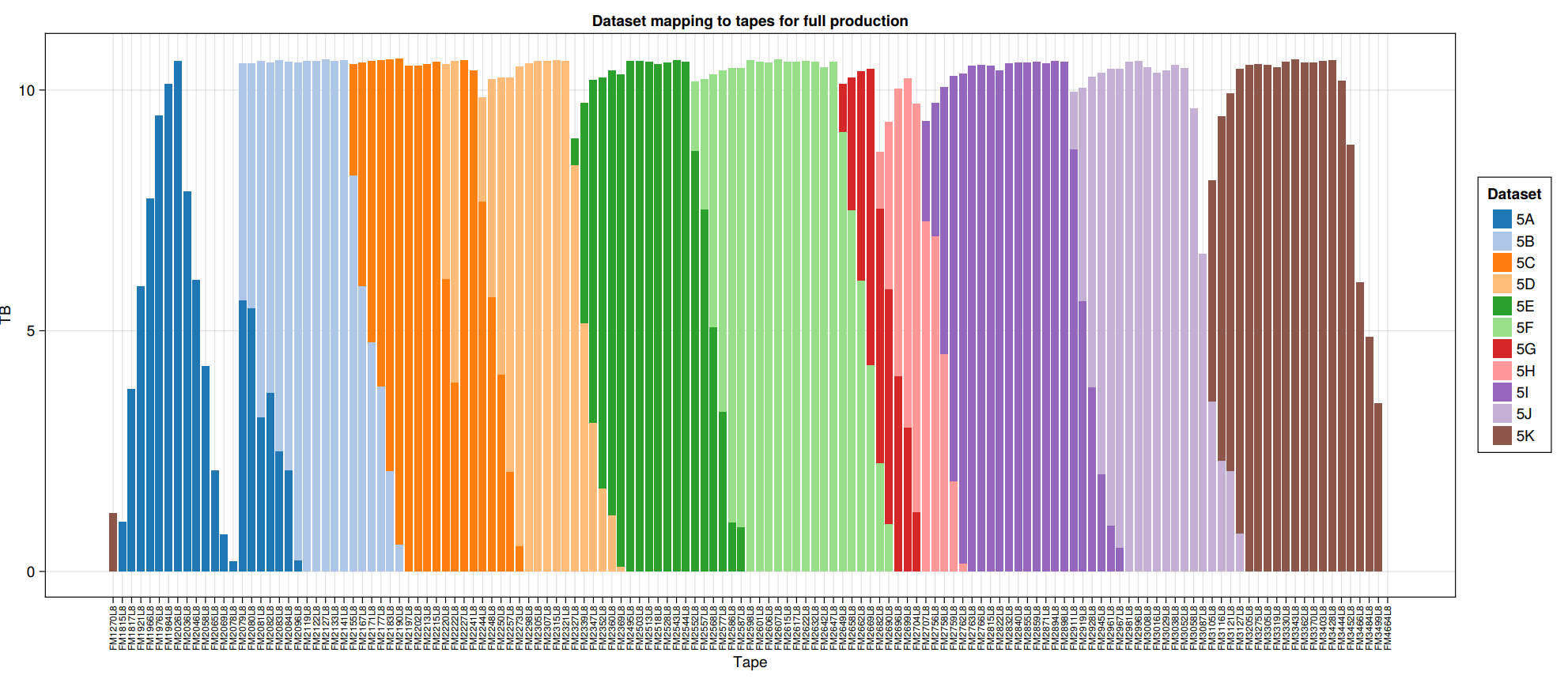}
    \caption{Example mapping of Run-5 full-production datasets onto tape volumes. Each vertical bar represents a tape volume, and the stacked colors show the amount of data from each dataset stored on that tape. The plot illustrates why dataset-level prestaging can be inefficient: a single dataset may be distributed over many tapes, while a single tape may contain files from multiple datasets. Migration mode reduces repeated tape mounts by staging complete selected tape volumes when most of their contents are needed.}
    \label{fig:datasets-on-tape}
\end{figure*}

Figure~\ref{fig:datasets-on-tape} illustrates the mapping of Run-5 full-production datasets onto tape volumes. It shows why dataset-level prestaging can be inefficient: individual datasets can span many tapes, and individual tapes can contain files from multiple datasets. This structure motivated the use of migration mode for campaigns in which most of the contents of the relevant tapes were eventually required.

\subsubsection{File Transfer}
The FTS is designed for file transfer between tape systems and disk locations. It monitors selected directories, usually those designated as output directories via the POMS tool~\cite{POMS_2020} during job submissions, for new files. It then catalogs and archives any discovered files to specified locations based on predefined rules set in a configuration file. Additionally, FTS supports customized plugins capable of extracting experiment-specific metadata. This metadata is then used to register associated files with the SAM system.

\subsubsection{Automation and Workflow Optimization}
Muon \gmtwo offline and simulation processing were largely automated using the POMS tool. POMS offers a web service for automated job submission to distributed resources as requested by the user. It also facilitated monitoring, recovery of failed submissions, debugging, and record-keeping. POMS supported the execution of multiple dependent stages, where a secondary stage starts only after the primary stage has completed and uses the output of the primary stage as its input. This feature has been used extensively in the \gmtwo experiment, enabling seamless progression between stages of data processing. This automation has reduced manual intervention significantly, allowing the production team to focus on high-level optimization rather than routine task execution.

Additionally, a dedicated POMS user profile called ``production-shifter'' allowed non-expert collaborators to assist the production team. This profile granted permissions to a selection of commands, enabling these collaborators to monitor and launch processing campaigns as needed. This enhancement streamlined the workflow and improved the efficiency of data processing in the \gmtwo experiment.

\section{Offline production workflow}\label{sec:production}

\begin{figure*}[!ht]
	\centering
	\includegraphics[width=0.9\textwidth]{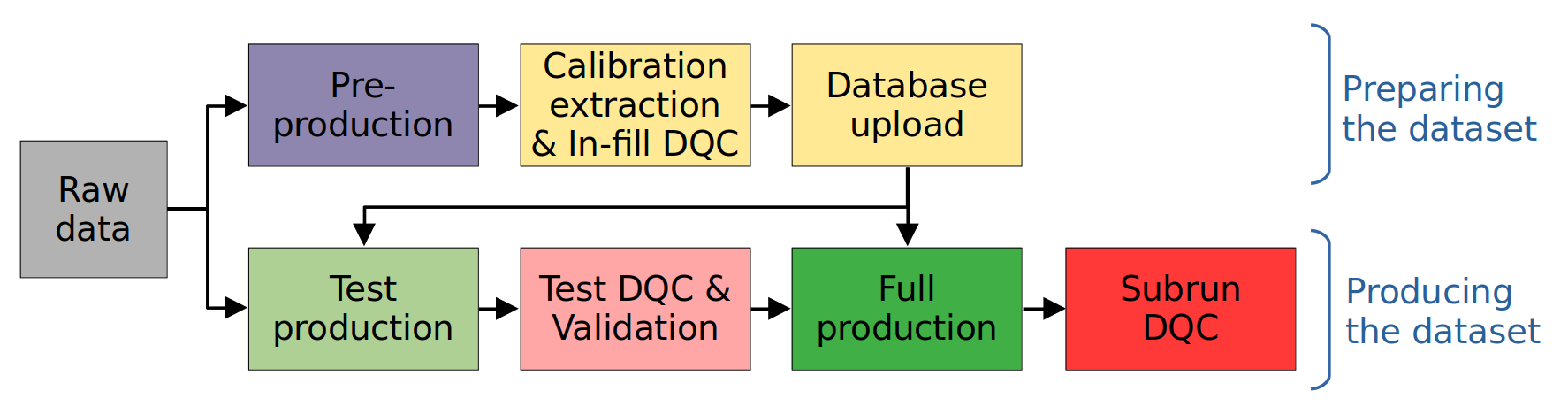}
    \caption{Offline production workflow of the Muon \gmtwo Experiment.}
	\label{fig:workflow}
\end{figure*}

The production of the experimental data was divided into two major phases. The first phase (``preproduction") prepared the dataset and consisted of a minimal reconstruction of the calorimeter data to extract calibration constants. The second phase (``production") produced the dataset by performing the complete reconstruction of the calorimeter and tracker data and applying the calibration constants determined in the first phase. Both phases started from the raw data and incorporated Data Quality Control (DQC) tests to ensure that the data satisfied nominal running conditions. The production workflow was broken down into 7 steps as illustrated in Figure \ref{fig:workflow}: \textit{Pre-production}, \textit{Calibration extraction and In-Fill DQC}, \textit{Database upload}, \textit{Test production}, \textit{Test DQC and Validation}, \textit{Full production}, and \textit{Subrun DQC}. To maximize resource usage and minimize overall production time, each step ran on at least one dataset at any given time. Each step generated one output file per input file so that failed files could be easily identified and reprocessed.

\subsection{Preparing the datasets}
\label{sec:pre-production}

The first step to prepare the dataset was to decompress the raw binary files and perform a minimal reconstruction of the calorimeter data. At this stage, only signal pulse fitting and energy calibration were performed. The waveform templates for the pulse fitting and the energy calibration constants were previously extracted in dedicated workflows on laser and beam data at the start of each running period.

The second step analyzed the reconstructed data to extract gain calibration constants and determine the quality of individual muon fills. A muon fill could be used for analysis if the data from all subsystems were within a nominal range. This step was highly automated and divided into independent tasks: one task for each subsystem.
These tasks ran in parallel and produced either json files or ROOT files with trees and histograms. These files were used by calibration experts to verify the results before uploading them to the database. 

Both of these steps were run with POMS, which executed the second step automatically after the first step completed. Dataset slices consisting of $O(1000)$ files were produced in each grid submission. A grid submission was made every 4 to 6 hours depending on grid node availability. A final recovery pass ran on the files that did not complete in the first submission. This resubmission was automatically handled by POMS.

The final step in this preparatory phase was to verify the extracted calibration constants were correct and to upload them to the \textit{conditions} database. Calibration experts manually ran the validation routines and the database upload. The calibration constants and quality flags were stored in \textit{data} and \textit{status} tables. The former contained the constants themselves, and the latter contained metadata and activation flags. Each table was finally marked with an interval of validity (IOV) stating the range of runs that the constants had been extracted from. The tables were uploaded to a development database first and, upon validation that no error occurred during upload, were then cloned to the production database. Finally, any raw files that failed to run through the pre-production process or that were found to contain no beam data were marked as such by updating its SAM metadata and did not proceed to full production.

\subsection{Producing the datasets} 

After the calibration constants were extracted and the good-quality files were identified, the dataset was ready for full production. This phase started from the raw data once again and performed a full reconstruction of all detectors, applying the calibration constants extracted during the data preparation phase. The full production involved the reconstruction of calorimeter data, tracker data, and auxiliary detectors. Four different reconstructions were performed on the same calorimeter data.

Before proceeding with the production of the entire dataset, a \textit{test} production was performed on a subset of it. A selection of $\sim$10\% of the files, uniformly distributed throughout the dataset, was processed first to detect potential reconstruction problems. A DQC step was performed on the produced files to ensure there were no errors or anomalies that needed further investigation. This validation step proved to be very effective in identifying software bugs and unexpected experimental conditions, as well as preventing a waste of grid resources on data that would be discarded or re-processed.

\begin{figure*}[!ht]
	\centering
    \includegraphics[width=\textwidth]{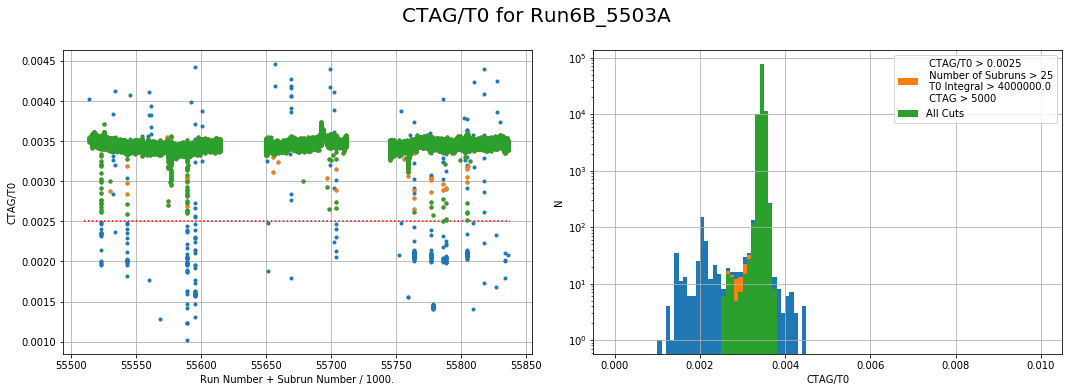}
    \caption{Example timeline generated during the DQC step showing the number of reconstructed positrons (``CTAGs") normalized to the muon beam intensity as a histogram (right) and as a function of file number (left). The red dashed line in the left-hand plot corresponds to the minimum value required for a file to pass this check. Blue points correspond to files that do not pass this check, orange points correspond to files that pass this check but do not pass other checks, and green points correspond to files that pass all checks in the DQC step. Only files represented by the green points are provided to analyzers for analysis.}
	\label{fig:dqc-example}
\end{figure*}

The DQC step was performed by shifters. The shifter checked two PDF booklet files: one containing gain validation histograms, and the other containing reconstruction validation histograms, as well as timelines of quantities such as the number of reconstructed positrons in each file. The shifter checked the PDF booklet files against reference booklets. The gain booklet contained information about laser-based gain calibrations for each calorimeter crystal, positron energy spectra for each calorimeter before and after gain corrections, and mean positron energies across the whole dataset. The reconstruction booklet plotted the positron energy and time spectra from the different reconstructions together to ensure consistency between the reconstructions. The shifter checked the timelines to ensure that quantities were stable. An example timeline is shown in Figure~\ref{fig:dqc-example}.

Once any reconstruction anomalies or timeline instabilities in the \textit{test} production were solved, full production resumed on the remainder of the dataset. When full production was complete, the DQC step was repeated with the results now uploaded to a Postgres database. This database contained a table for each DQC check with each row representing an individual file. Finally, the list of files that passed all DQC categories was extracted and the metadata flags on the raw MIDAS files were updated in SAM. These metadata flags were then used to define the final datasets delivered to analyzers.  \\

\section{Computing workflow for analysis}\label{sec:analysis}

The output of the data production described in Section \ref{sec:production} were \textit{art-root} files containing the reconstructed data in the form of \textit{art data products}. While the format of these files was convenient for preserving the history of the reconstruction steps and the associated quality filters, they presented some challenges to the analyzers:
\begin{itemize}
    \item Large number of files: because each data file was processed separately, the total number of files that passed all data quality checks was roughly 3.6 million for Run-1 to Run-6. Analyzers needed to analyze every single file in order to achieve the experiment's precision goal. The handling of such a large number of files posed a challenge. For example, if an analyzer's grid job failed, they needed to identify which files needed processing again.
    \item Very large dataset size: each output file contained the complete reconstruction chain of both calorimeter and tracker data, as well as auxiliary detectors. The complete reconstruction chain allowed analyzers to retrieve partially-reconstructed events to conduct systematic studies. However, this brought the total size of Run-1 to Run-6 reconstructed data to almost 9 PB.
    \item Availability: the size of the reconstructed data was too large to keep them staged on disk at all times. The analysis teams required frequent and on-demand access to the data in order to iterate their analysis algorithms and to conduct systematic studies. The size of Run-1 (0.3 PB) and Run-2/3 (1.7 PB) reconstructed data was small enough that they could be staged on the Fermilab general cache pool. However, this was not the case for the Run-4 to Run-6 analysis (6.6 PB).
    \item Portability: any analysis to be conducted on the \textit{art-root} files required the \textit{art} framework, which is a powerful tool but with a steep learning curve. Many analyzers preferred to handle their data using standard \texttt{ROOT} files that can be easily processed and inspected on personal computers.
\end{itemize}

To address these issues, most analyzers extracted their information of interest into a collection of \texttt{ROOT} files containing trees and histograms and run their analyses on those. As the data sets grew from Run-1 to Run-6, the production team gradually took care of creating these data-reduced ``\textit{analysis files}" for the analyzers.

\subsection{Art skims}

While \texttt{ROOT} files proved to be an efficient solution for the analyzers, access to the full offline files was sometimes needed to rerun full analysis routines or perform novel ones. As mentioned, the Run-4/5/6 dataset was too large to be stored on disk and be accessible at all times. A solution was found in creating a reduced version of the \textit{art} files, named \lq\lq art-skims\rq\rq, by removing all the data products that were not needed for the typical analysis.\\

For this purpose, only the uncalibrated and fully-calibrated calorimeter hits are kept, which achieved a five-fold size reduction. A further reduction could have been achieved by dropping the uncalibrated hits, but their presence ensured that custom gain calibrations can be applied for systematic studies. The disk cache needed to store Run-4/5/6 therefore reduced from 6.6 PB to 1.3 PB.\\

To reduce the number of overall files as well, the reduced \textit{art-skims} went through a merging process. Each final file contained 10 random files belonging to the same run number. We kept track of which files were merged as metadata.\\

When analyzers needed production data that was not part of the permanently cached datasets, they submitted a request to the production managers who prestaged the required data on demand. Depending on the available resources, these operations could take between a few hours and a few days to complete.

\subsection{Analysis Files}\label{sec:analysis_skims}

Toward the end of the experiment, five different types of analysis files were managed by the production team: two for the tracker data, and three for the calorimeter data. These contained the ROOT trees or ROOT histograms that analyzers needed for their analyses. For all of them, submissions were executed with POMS and the output was assigned with metadata with the same tools as the offline production.\\

The production was always conducted by generating one output file per input file. Therefore, an extra step was needed to merge the output to reduce the overall number of files that the analyzers needed to handle. As the analysis files were always ROOT files, the built-in \texttt{hadd} tool was used to merge them.\\

The merging procedure was performed in two or three steps, reducing the number of files by some orders of magnitude each time. The merging routines were performed either on the FermiGrid with POMS submissions or on local machines. Considering a typical dataset contained 100,000 files, the first step merged the files into groups of 100 on the grid. This way, the job could be parallelized over 1000 cores. The final step, consisting of merging the 1000 intermediate files into one, was performed locally on a server with a high core count. This way, the single \texttt{hadd} job could be run multi-threaded. While the intermediate files were stored on scratch disks, the final results were stored either on tape or persistent disks.

\subsection{Computing resources for analyzers}

The analysis of Muon \gmtwo data was performed by the analyzers over multiple Fermilab-owned servers or on resources locally owned by the participating institutions. The collaboration-wide resources made available by Fermilab consisted of eight virtual machines and one high-performance workstation. The virtual machines had four CPU threads and 12 GB of memory each. Each user was given 400 GB disk space for storing personal data. The workstation hosted 48 CPU threads, 187 GB of memory, and 7 TB of local disk space. This workstation was used for running local file merging jobs during production, as described in Section \ref{sec:analysis_skims}. These were characterized by having high-count CPU threads, large memory, and a combination of large disks and network-attached storage (NAS) servers for data storage.\\

CPU-intensive analysis jobs were run on FermiGrid and OSG, using the same resources shared by the Muon \gmtwo production. In addition to the reserved grid quota for production, there was always some overhead that allowed analyzers to run their jobs. The submission scripts were the same as for official production, and they provided failure recovery routines. They were generally executed manually via the virtual machines, with the exception of a few cases where POMS was used for analysis jobs.\\

While institution-owned NAS servers proved to be very useful in holding skimmed datasets for analysis, they presented a bottleneck when CPU intensive jobs had to be executed over the data stored there. The grid nodes can only interface to Fermilab central disk cache for I/O, so any NAS data required as input for the grid jobs had to be temporarily copied to a scratch cache folder on the corresponding node first. Because the NAS servers were not within the faster FermiGrid Network, copy speeds were limited to 10 Gbps, with effective speeds as low as 100 MB/s when the network is busy. As the size of a typical analyzed dataset is on the order of 10 TB, the required copy could take a few days.\\

The workflow schemes described in this section allowed all analyzers, despite having different requirements, to efficiently run their analyses. They ran on a reduced number of files and could access the data at any time without needing to prestage from tape. They could also use the production grid submission scripts for their own data processing needs and had access to dedicated machines to analyze the data. These procedures were developed for the $a_{\mu}$ analysis but they are also being used for the EDM, CPTLV, and Dark Matter analyses on the experiment.


\section{Simulation\label{sec:simulation}}

The Muon \gmtwo Experiment simulated the muon beam in the storage ring with several simulation packages. These simulations were used to optimize the experiment's design and to understand the effect of beam dynamics on the final analysis. The most computationally-intensive simulation involved injecting muons into the storage region and tracking them for over $200\,$km as they go around the storage ring 4700 times in a single \SI{700}{\micro\second} fill. Accurately tracking single muons in the simulation over this distance placed high demands on the tracking tools and required symplectic integrators that strictly enforced energy conservation throughout the entire path. Depending on the simulation package, these tasks could require long CPU running times resulting in $\sim$500 seconds running time for each muon that was stored for the entire fill. All of these considerations meant it was not feasible to produce or store simulated samples that were the same size as the $\sim$10$^{11}$ muon decay data samples collected during the experiment's lifetime. Instead, a variety of tools and techniques were optimized and deployed to answer the questions at hand. This Section will briefly introduce the various simulation packages, describe the primary application for each tool and document the computing resources needed to produce the simulation samples. An introduction to the muon campus, storage ring components, simulation packages and associated beam dynamics effects may be found in the first Muon \gmtwo beam dynamics paper~\cite{Muong-2:2021xzz}.

\subsection{Simulation Packages}

In the experiment, muons were produced from the decay of the pions that were generated by scattering protons on an iron-nickel-chromium target. This process was simulated with two packages, \texttt{MARS}\cite{MARS}\cite{Mokhov:2014soa} and \texttt{G4Beamline}\cite{G4Beamline}. The redundancy allowed for the vetting of the model implementation and the underlying physics in each program. The results from both packages were in turn benchmarked against Fermilab test beam data. The resulting positively charged pions, muons, positrons, and protons that emerge from the downstream exit of the lithium focusing lens were saved for further tracking in the \texttt{G4Beamline} model of the delivery ring and the \texttt{BMAD}\cite{SAGAN2006356} model of the final focus~\cite{Muong-2:2024hpx}. The primary goal of this combination of programs was to produce distributions of highly polarized $\mu^+$ that can serve as input to the packages that simulate muon injection and storage in the ring.

The time-dependent electromagnetic fields present in the storage ring were modeled by three packages, \texttt{BMAD}, \texttt{COSY INFINITY}\cite{Berz:2026eck, 10.1093/jmicro/dfag019} and \texttt{GM2Ringsim}~\cite{PhysRevAccelBeams.24.044002}\cite{Muong-2:2024hpx}. The latter of these was based on Geant4~\cite{ALLISON2016186,1610988,AGOSTINELLI2003250}. Furthermore, \texttt{BMAD} and \texttt{GM2Ringsim} implemented an injection model, including the orientation of the beam as it exits the final focus beamline and enters the storage ring. 

\begin{figure*}[!ht]
\centering	\includegraphics[width=\textwidth]{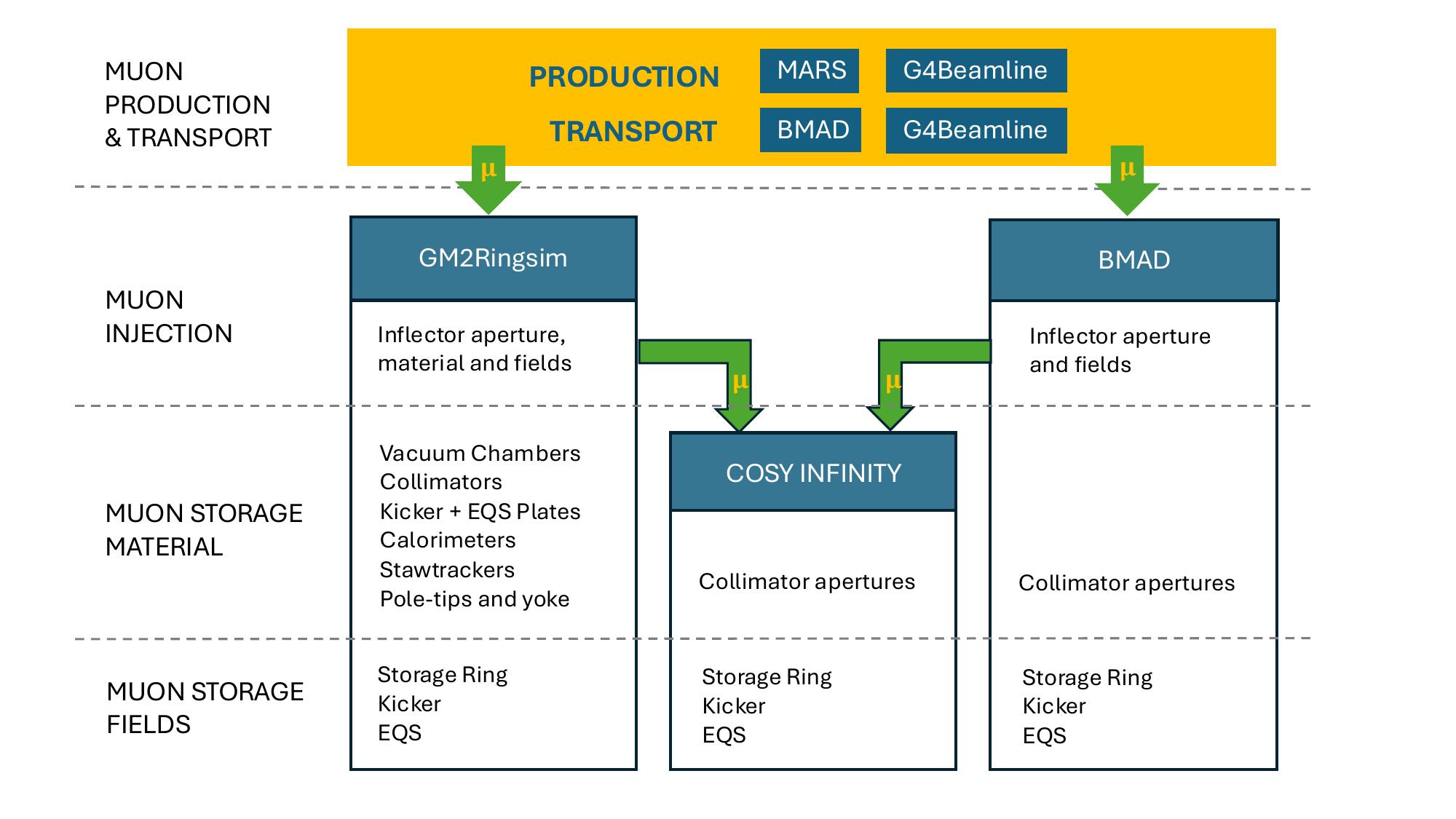}\caption{This diagram shows the stages of the Muon \gmtwo simulation infrastructure and the relationship between the packages discussed in Section \ref{sec:simulation}. Either \texttt{MARS} or \texttt{G4Beamline} is used to simulate particle production at the target, followed by transport through the beamlines and delivery ring via a combinations of \texttt{BMAD} and \texttt{G4Beamline}. The output of these packages is the phase-space distribution of the muons at the end of the M5 line, which in turn serves as input to injection through the inflector in the \texttt{BMAD} and \texttt{GM2Ringsim} implementations of the storage ring. The \texttt{COSY INFINITY} package takes the muon distributions evaluated at the end of the inflector in \texttt{GM2Ringsim} or \texttt{BMAD} as input.}
\label{fig:SIMUFlow}
\end{figure*}

The graphic in Figure \ref{fig:SIMUFlow} shows the simulation data production stages for the Muon \gmtwo experiment and the programs used to generate these data. The blue blocks represent packages that contribute to each stage and the green arrows represent the muon distributions that result from each stage and are subsequently fed into a new stage. \texttt{COSY INFINITY} did not include an injection model so the input muon distributions were extracted at the inflector exit from either \texttt{BMAD} or \texttt{GM2Ringsim}. Once inside the storage ring, each package has the ability to track the location, momentum, and spin of the muons at specific times around the ring. The \texttt{GM2Ringsim} package can also track muon decay products and particles produced in material interactions. Although the three packages all used the same field maps as input, there were important differences in how muon tracking is performed. For example, \texttt{GM2Ringsim} is the only package that fully accounts for material effects, while \texttt{COSY INFINITY} is capable of using fully symplectic integration. The complementarity of these tools provided essential consistency checks on the extracted physics observables.

\subsection{Simulation Production and Workflow}

\subsubsection{\texttt{MARS} simulations of the target station} The \texttt{MARS} studies were initially run at the RHIC ATLAS Computing Facility and then later on FermiGrid. The \texttt{MARS} package was used to produce $\sim$$10^9$ protons-on-target (POT) over the course of several days. Note this number is three orders of magnitude less than the typical POT per muon fill during the experiment and a subsequent implementation in \texttt{G4Beamline} was used to produce $\sim$$10^{12}$ POT samples. The storage requirements for \texttt{MARS} were low, only 0.2 GB of disk space needed for a 1B POT sample. The output was stored in a format compatible with other tools for input. 

\subsubsection{\texttt{G4beamline} simulations of the target station and particle transport beamlines} A \texttt{G4beamline} Docker image was used to run at NERSC as well as OSG. 
Several days to a week on the OSG were needed to simulate $\sim$$10^{12}$ POT, depending on the availability of computing resources. The generation of particle distributions at the point of injection into the delivery ring required about 1.8 GB per $\sim$$10^{12}$ POT for muons and pions. 

\subsubsection{\texttt{BMAD}}
\texttt{BMAD} simulations were run at several sites. For the delivery ring and beamline simulations, \texttt{BMAD} was compiled and run on either a local laptop or a local workstation (16 core, 32 thread processor). Typically, a run of $3\times10^{12}$ POT completed within 24 hours and required $\sim300$ MB of storage space once converted to \texttt{G4beamline}'s BLTrackFile2 format. 

The \texttt{BMAD} simulations of the injection line and storage ring were typically run on the CLASSE computing grid at Cornell University. The typical running time there was about 12 hours, with an output size on the order of a GB stored on a local pair of Tier-1 dedicated disks. The output of the simulation was stored in text files and the data was processed with some combination of custom Fortran programs and gnuplot.

\subsubsection{\texttt{COSY INFINITY}}
\texttt{COSY INFINITY} productions were run at the Michigan State University computing clusters and on the Cori supercomputer at the NERSC. Beam dynamics corrections and magnetic field convolutions required about 2 days per experimental setting, one day to prepare the non-linear transfer maps and another day for beam tracking. About one TB of disk space was required for map and beam-distribution storage.

\subsubsection{\texttt{GM2Ringsim}}
The \texttt{GM2Ringsim} package was typically run in two different modes. The first, called beamgun mode, injected a distribution of muons and tracks them for a finite time around the ring. Particle interactions with the detectors, as well as their position, momentum, and spin were recorded in \textit{ghost tracking planes}, which are virtual detectors placed in strategic locations on the inside of the ring. These ghost tracking planes used the parallel world functionality in GEANT4 \cite{ALLISON2016186,1610988,AGOSTINELLI2003250}, which allowed users to extract information about a particle at a specific location without imposing any effects from material interactions. Often it was more efficient to artificially extend the lifetime of the circulating motions as well, for example to study the time dependence of radial and vertical positions and betatron oscillations. Typically a 3~M event sample was produced with decay turned off for each set of experimental conditions. Unlike the other simulation packages, production of \texttt{GM2Ringsim} beam gun simulation data closely followed the production of the experimental data. Jobs were submitted using POMS to FermiGrid and the OSG. The simulation output was stored in art-ROOT files. The output files were transferred back to Fermilab and then to tape using FTS. Datasets were logged in the SAM dataset access system. In the case of the second publication from the Muon \gmtwo experiment (Runs 2 + 3) thirty simulated datasets were produced to study beam dynamics corrections and to estimate systematic errors. Each dataset took approximately three days to produce, using the 2000 slots allocated by Fermilab for the \gmtwo simulation production team. The storage output for each dataset was on the order of 5~TB. A summary of the simulation production is give in Table~\ref{tab:simproduction}.

The second mode, called gas-gun mode, skipped injection completely and instead populated the ring with muons that decay immediately to positrons that are then tracked through the electromagnetic fields and into the calorimeter and straw tracker detectors that were used to reconstruct the positron energy and muon decay location. This mode allowed for efficient simulation production for studies that focused on the interaction of the decay positrons with the detectors, for example acceptance effects and any correlations with spin and momentum. Since the positrons traverse less than half of the ring before exiting, the time needed to track each event is orders of magnitude less when running in gas-gun mode. That said, acceptance studies required a much higher fidelity than the beam dynamics studies that used the beam gun. A 10~B event gas-gun simulation sample was produced to study the phase acceptance correction and systematic error for the analysis. The sample was produced at NERSC, which was chosen for its high throughput. Jobs were launched from both FermiGrid and NERSC. A test job on one 68 core NERSC KNL node took 100 minutes to produce 5.4~M events. The output was again stored in \textit{art} data products and in ROOT files totaling 0.8~PB. It took several months to complete the entire 10 billion events. The data was temporarily stored on NERSC's Cori system as it was produced and then transfered to FermiGrid via a specialized FTS protocol.


\begin{table*}[!ht]
\centering
\begin{tabular}{c|c|c|c} 
Run Period & Gun & $\#$Events (M) & Data Volume (PB)\\\hline\hline
Run 1 & gas & 10,800 & 0.8\\
Run 1 & beam & 1330 & 2.21 \\
Run 2 & beam & 1066 & 1.78 \\
Run 3 & beam & 78 & 0.12 \\
Run 4 & beam & 4609 & 0.69 \\
Run 5/6 & beam & 122 & 0.17 \\
\end{tabular}
\caption{Table of simulation dataset sizes for each running period. Note that Run 5 and Run 6 had the same experimental configuration and so a common simulation dataset was produced.}
\label{tab:simproduction}
\end{table*}

\section{Timeline and improvements}\label{sec:timeline}

The Muon \gmtwo Experiment at Fermilab collected physics data from 2018 to 2023. By the end of data taking, production was processing datasets as they were collected. To get to this position we made many improvements over the years, which we will outline in this section.

Offline reconstruction began slowly after Run-1, as the software was still being finalized and the data was used to refine the process. Consequently, the Run-1 datasets were reprocessed multiple times — Run-1A alone was processed eleven times before reaching publication-ready form. 
Additionally, the offline production workflow was not optimized to process large amounts of data efficiently. The production turnaround time, i.e. the time elapsed between the end of beam data taking and production completion, was 1.5 years as shown in Table \ref{tab:production}.\\
The collected statistics and data size then increased by a factor of five from \RunOne to Run-2/3 and then an additional factor of four from Run-2/3 to Run-4/5/6. The experiment, together with the support from the Fermilab computing division, managed to catch up with the data processing backlog after \RunOne and then kept pace with the ever-increasing amount of data by continuously improving on the efficiency of the production processes, the production data rate capability, and the production management.

\begin{figure*}[!ht]
    \centering \includegraphics[width=0.72\textwidth]{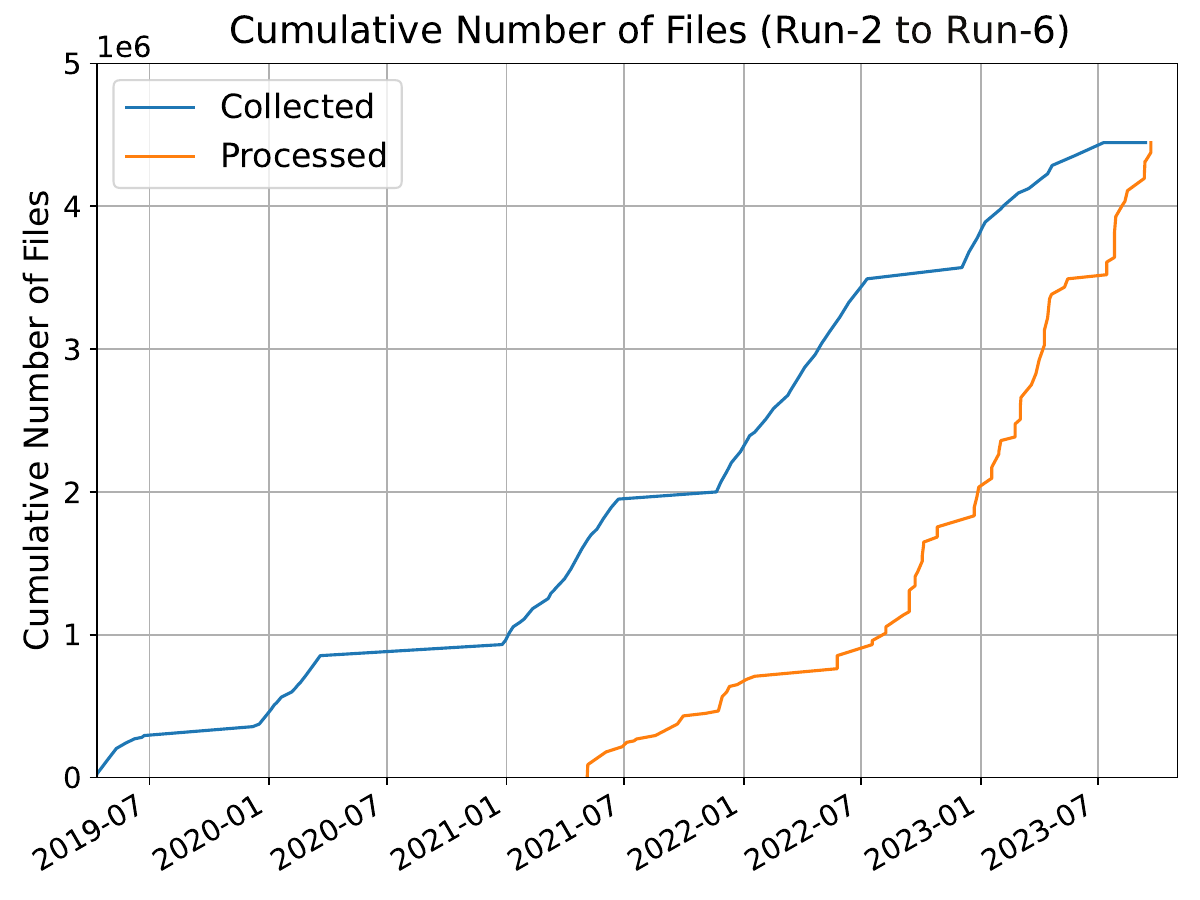}
    \caption{
    Cumulative number of raw data files collected by the DAQ (blue) and cumulative number of files processed through the offline production workflow and used in the final analyses (orange) from Run-2 to Run-6. The vertical separation between the curves represents the processing backlog at a given time, while their convergence near the end of the experiment shows that production caught up with data collection.}
    \label{fig:timeline}
\end{figure*}

After \RunOne reconstruction was finalized and ready for analysis, a multi-phase production workflow was developed for \RunTwo production, as described in Section \ref{sec:production}. At the same time, the production management tool POMS and the conditions database described in Section \ref{sec:comp_infrastructure} were implemented. Before the implementation of the database, the \RunOne calibration constants were stored in text files. The time to implement these software and workflow renovations further delayed the production of the \RunTwo and \RunThree datasets. However, the new procedures allowed for further optimization, parallelization, and automation of the processes, consequently speeding up the production rate by an order of magnitude overall. Despite the much larger size of \RunFour, \RunFive, and \RunSix, together contributing to $\sim$75\% of the total statistics and size, their production turnaround time was continuously reduced to the final 2.5 months for \RunSix. Figure \ref{fig:timeline} shows the data production timeline across the years.\\

The most impactful improvements were:
\begin{itemize}
    \item \textbf{Computing resources.}
        \subitem{+} The number of slots in Fermigrid reserved to \gmtwo was increased from $2500$ to $5700$ slots, with each slot representing 1 CPU thread and 2 GB of memory. This allowed us to run more jobs in parallel. 
        \subitem{+} The memory footprint of the production jobs was decreased to be less than 2 GB, so that each would occupy only one grid slot. This required optimizations in both the production code and the \textit{art} job configurations.
        \subitem{+} The number of files processed per grid submission was increased from 5000 to 20000, and the number of files processed per job was increased from 2 to 8. This increased the grid occupancy by reducing the amount of time passing between the processing of consecutive slices of 5000 files. In fact, the time it took for a job to process any particular file strongly depended on both the number of stored muons and the specific CPU model where the job ended up running. As a consequence, while some jobs might end earlier, some others could take up to twice the time to run in any specific slice of 5000 files. Increasing the slice size helped take advantage of the faster nodes and increase the overall efficiency.
    
    \item \textbf{Data handling.}
        \subitem{+} The implementation of the conditions database, optimized with an HTTPS interface for multiple concurrent requests, streamlined the management of data calibration and simulation constants.
        \subitem{+} The introduction of ``migration mode", described in Section \ref{sec:comp_infrastructure}, increased the prestaging speed by a factor of 3 to 5, effectively addressing one of the largest bottlenecks in offline production by enabling faster data transfer from tape to disk storage.
        \subitem{+} Optimal dataset size was found to be 100,000 files. This change leveraged the larger processing slices of 20,000 and minimized the number of steps that had to be repeated for each dataset, such as recoveries and validation routines. In principle, larger datasets, e.g. totaling 200k files, would have been even more efficient. However, some processes, such as the histogram merging, would have then exceeded the local resources available.
        \subitem{+} During \RunSix data taking the pre-production phase ran automatically. This was done when the new raw files coming from DAQ were stored on the temporary disk cache pool while waiting to get copied to tape. The \gmtwo data-taking speed was roughly 21 TB/day and the cache pool capacity was 114 TB, so newly acquired raw files were cached for five days before making space for the new incoming files. Each day, automatic jobs processed $\sim10^4$ raw files. The few files that failed this automated production step, typically $\sim$5$\%$, were staged and reprocessed manually on a weekly basis. This automated process sped up the pre-production phase significantly and allowed for the \RunSix full-production to be completed promptly after data collection. 
    
    \item \textbf{Management and human resources.}
        \subitem{+} Validation routines at the end of the full-production step and the introduction of the test production greatly helped uncover calibration errors and algorithm bugs. This substantially reduced the number of times that a given dataset would be produced in order to reach the final production.
        \subitem{+} The entire collaboration participated in offline production shifts during the production of Run 3 in November 2021. These shifts counted towards collaborators' shift credits. The preparation for getting the shifts ready required documenting the entire production process and facilitated the formalization of the core processes and the clean up of the unnecessary steps. Though time-consuming, this greatly benefited subsequent workflow optimization. The shifters schedule ensured that the production could run at maximum speed at all times, without being limited by the time availability of the few production experts.
        \subitem{+} Technical and resource support from Fermilab. Infrastructure bottlenecks were addressed individually by the infrastructure expert teams, most notably raw data cache availability, resolved via a dedicated cache pool and new prestaging mode. Computing priority was granted during peak demand to meet publication deadlines.
                 
\end{itemize}

Thanks to these and other optimizations, the full production overall speed improved by a factor of two between \RunTwo and \RunThree productions, and by another factor of two between \RunThree and \RunFour productions, as demonstrated in Table \ref{tab:production}. The slightly slower \RunSix rate shown in the last row indicates that the production speed was sufficient to keep up with the experiment data taking rate.\\

\begin{table*}[!ht]
	\centering
	\begin{tabular}{c|c|c|c|c|c}
		Dataset&N° of files&Size [TB]&Production time [days]&Rate [files/day]&Days after run ended\\\hline
        \RunOne&258108&544&572&451&565\\
		\RunTwo&267152&563&129&2071&750\\
		\RunThree&538622&1066&122&4415&794\\
		\RunFour&992005&1953&111&8937&550\\
        \RunFive&1453853&2838&141&10311&305\\
        \RunSix&748724&1760&106&7063&74\\\hline
	\end{tabular}
	\caption{Production turnaround time per aggregated dataset. Except for Run-1, where the pre-production phase was not established yet, only the full-production processing time has been considered.}
	\label{tab:production}
\end{table*}

\section{Conclusions}\label{sec:conclusion}

We presented lessons learned from developing and operating the computing infrastructure for the Muon \gmtwo Experiment, which may benefit future medium-sized experiments collecting data at the Petabyte scale.

\begin{itemize}
    \item \textbf{Software and computing}
        \subitem{+} Data availability at the multi-Petabyte scale has always been one of the main bottlenecks of Muon \gmtwo production. A lot of effort went into easing the process of staging files from tape to disk. A dedicated cache pool and tape-based staging routine were implemented, complemented by collaboration-developed data staging schedules and monitoring tools.
    
        \subitem{+} Maintaining diverse, scalable computing resources proved essential. The use of off-site resources as OpenScienceGrid nodes was very helpful in times of high loads. Part of the simulation jobs were executed at the High Performance Computing (HPC) facility NERSC and tracker reconstruction was offloaded to UK resources.

        \subitem{+} Software readiness, including robust calibration infrastructure, is crucial for timely data processing. The Muon g-2 experiment initially lacked a conditions database for its large volume of calibration constants; implementing one later required months of refactoring. While some fine-tuning with real data is inevitable, core infrastructure should be in place before data-taking begins.
    
    \item \textbf{Management}
        \subitem{+} Software validation is crucial to avoid reprocessing and idle times. The highly modular structure of \textit{art} code, useful for HEP reconstructions, allows independent people to work separately on various pieces of the reconstruction chain. However, a lack of communication between developers and no clear code review procedures in earlier times often resulted in subtle bugs going unnoticed and entering production, only to be found out at the analysis stage. The addition of validation steps throughout the production process, which performed physics sanity checks and calibration cross-checks, increased the bug-spotting efficiency and dramatically decreased the amount of reprocessing required. A stronger code review system would have been beneficial.
        
        \subitem{+} Offline computing was a collaboration-wide effort and, when running production at full speed needed ~10 FTE to run at a steady state. The introduction of offline shifts helped keep the production running at all times. As the workflow was fine-tuned, further parallelization and automation of the process decreased the number of shifters from 8 to 3.
        
        \subitem{+} Documentation, instructions, and bookkeeping are essential. Sometimes a five-year-old file had to be recreated from scratch due to data corruption or similar issues. The exact same production process had to be easily replicated in the same way and by different people. \textit{POMS} was an excellent tool for bookkeeping and production management.
    
        \subitem{+} Support from the hosting laboratory on infrastructure and expertise has been crucial for the initial development of the experiment computing model and for the sustained operation. A computing liaison between the collaboration and the laboratory proved to be an important role. This person helped organize crisis response teams when major infrastructure bottlenecks or service interruptions occurred.
        
    
\end{itemize}

The computing model described in this paper, the involvement of the entire collaboration, and the support of the hosting laboratory were fundamental to the success of the Muon \gmtwo Experiment. With approximately 20 PB of raw, reconstructed, and simulated data, and more than 150 million CPU hours of distributed computing, the Muon \gmtwo collaboration transformed about 150 billion recorded muon decays into the most precise determination of the muon anomalous magnetic moment.

\section*{Acknowledgements}

We thank the Fermilab management and staff, including many computing professionals, for their strong support of this experiment,
as well as our university and national laboratory engineers, technicians, and workshops for their tremendous support. 

The Muon \gmtwo\ Experiment was performed at the Fermi National Accelerator Laboratory, a U.S. Department of Energy, Office of Science, HEP User Facility. Fermilab is managed by Fermi Forward Discovery Group, LLC, acting under Contract No. 89243024CSC000002. Additional support for the experiment was provided by the U.S. DOE Office of Science through the offices of HEP, NP, and ASCR, by the U.S.-Japan Science and Technology Cooperation Program in HEP, by the National Science Foundation (USA), by the Istituto Nazionale di Fisica Nucleare (Italy), by the Science and Technology Facilities Council (UK) and the Royal Society (UK) under Grant No.~URF$\backslash$R1$\backslash$231503, 
by the National Natural Science Foundation of China (Grant No. 12475108, 12305217, 12075151), by the Ministry of Science, ICT and Future Planning (MSIP), the National Research Foundation of Korea (NRF), and the Institute for Basic Science under Project Code  No.~IBS-R017-D1 (Republic of Korea), by the German Research Foundation (DFG) 
through the Cluster of Excellence PRISMA+ (EXC 2118/1, Project ID 39083149) and the 
Cluster of Excellence PRISMA++ (EXC 2118/2, Project ID 39083149), by the European Union 
Horizon 2020 research and innovation program under the Marie Sk\l{}odowska-Curie Grant Agreements No. 101006726 and No. 734303, by the European Union STRONG 2020 project under Grant Agreement No. 824093, and by the Leverhulme Trust, LIP-2021-014.

\bibliography{references}

\end{document}